\documentstyle[preprint,epsfig,eqsecnum,aps]{revtex}

\newcommand{\be}{\begin{equation}}
\newcommand{\ee}{\end{equation}}
\newcommand{\bea}{\begin{eqnarray}}
\newcommand{\eea}{\end{eqnarray}}
\begin{document}
\draft
\title{New features of some proton-neutron collective states}
\author{A. A. Raduta$^{a,b,c)}$, C. M. Raduta$^{c)}$, B. Codirla$^{b,d)}$,}
\address{$^{a)}$Instituto de Estructura de la Materia, CSIC, Serrano
 119-123, 28006 Madrid, Spain}
\address{$^{b)}$Dep. of Theoretical Physics and
Mathematics, Bucharest University, POB MG11, Romania}
\address{$^{c)}$Institute of Physics and Nuclear Engineering, Bucharest, 
POBox MG6, Romania}
\address{$^{d)}$Institut fur Theoretische Physik der Universitaet
Tuebingen, auf der Morgenstelle 14, Germany}
\date{\today}
\maketitle
\begin{abstract}
Using a schematic solvable many-body Hamiltonian, one studies a new type of
proton-neutron excitations within a time dependent variational approach.
Classical equations of motion are linearized  and subsequently
solved analytically. 
The harmonic state energy is compared with the energy of the first 
excited state provided by diagonalization as well as with the energies obtained by a
renormalized RPA and a boson expansion procedure.
The new collective mode describes a wobbling motion, in the space of isospin, 
and collapses for a  particle-particle interaction strength which is much larger than the physical value. 
A suggestion for the description 
of the system in the second nuclear phase is made. We identified the transition operators which might excite the new mode
from the ground state. 
\end{abstract}
\narrowtext
\section{Introduction}
\label{sec:level1}
One of the most exciting subject in theoretical nuclear physics is the
double beta decay, especially due to the neutrino-less ($0\nu\beta\beta$)
process
\cite{Hax,Fas1,Fas2,Suh}. Indeed, its discovery would answer a 
fundamental question whether
neutrino is a Majorana or a Dirac particle. The theories devoted to the 
description of this process suffer of the lack of reliable tests for the
nuclear matrix elements. O possibility to overcome such difficulties would
be to use the matrix elements which describe realistically the rate of
$2\nu\beta\beta$ decay. In this context many theoretical work have been
focussed on $2\nu\beta\beta$ process. Most formalisms are based on the
proton-neutron quasiparticle random phase approximation (pnQRPA) 
which includes the particle-particle ($pp$) channel in the two body interaction.
Since such an interaction is not considered in the mean field equations the
approach fails at a critical value of the interaction strength, $g_{pp}$.
Before this value is reached, the Gamow-Teller transition amplitude
($M_{GT}$) is decreasing rapidly and after a short interval is 
becoming equal to
zero. The experimental data for this amplitude is reached for a value of
$g_{pp}$ close to that one which vanishes $M_{GT}$ and also close to
the critical value.
Along the time, the instability of the pnQRPA ground state was considered
in different approaches. The first formalism devoted to this feature 
includes anharmonicities through the boson expansion technique
\cite{Rad1,Rad2,Suh1,Grif}.
Another method is the renormalized pnQRPA procedure (pnQRRPA)
\cite{Toi} which keeps
the harmonic picture but the actual boson is renormalized by effects coming
from the terms of the commutators algebra, which are not taken into
account in the standard pnQRPA approach.

In a previous paper\cite{Rad3} we have proved that the pnQRRPA procedure does not
include the additional effects in an consistent way. Indeed, if the 
commutators of two quasiparticle operators involves the average of
monopole terms then these terms should be considered also in the
commutators of the scattering terms. If one does so, new degrees of freedom
are switched on and a new pnQRPA boson can be defined. This contains,
besides the standard two quasiparticle operators, the proton-neutron
quasiparticles scattering terms. If the amplitude of the scattering term
is dominant comparing it to the  other amplitudes, the
pnQRRPA phonon describes a new nuclear state.

The aim of this paper is to show that such a mode appears in a natural way
within a time dependent treatment. The present approach points out new
properties of the new proton-neutron collective mode. We use a schematic
many-body Hamiltonian which for a single j-shell is exactly solvable.
In this way the approximations might be judged by comparing the predictions of the actual model with the 
corresponding exact results. Since the semi-classical treatment is the proper way
to determine the mean field, one expects that the present approach
is suitable to account for ground state correlations in a consistent way
 and therefore some of the
drawbacks mentioned in a previous publication \cite{Rad3}, like the breaking down
of the fully renormalized RPA before the standard RPA breaks down, are removed.
To understand better the virtues of the present model we compare
its predictions with the results obtained in a renormalized RPA approach and
a boson expansion formalism. Since the semi-classical methods have, sometimes, 
intuitive grounds we aim at obtaining a clear interpretation for the new
proton-neutron mode.  It is well known that the breaking down of the RPA
approach is associated to a phase transition. In this respect the semi-classical 
formalism is a suitable framework to define the nuclear phases which are bridged
by the Goldstone mode. Above arguments justify our option for a semi-classical treatment
and also sketch a set of expectations.

This project is achieved according to the following plan:
In Section II, we describe the model Hamiltonian. The main features of the 
fully renormalized RPA approach, presented in a previous paper, are
briefly reviewed. A time dependent variational principle is formulated
in connection with a truncated quasiparticle Hamiltonian, in Section III. This Hamiltonian
is the term of the model Hamiltonian which determines the equations of
motion for the quasiparticle proton-neutron scattering terms, in the
de-coupling regime. The classical equations of motion
and their solutions are presented in Section IV. 
The new $pn$ collective mode is 
alternatively described through the renormalized RPA approach and boson expansion formalism
in Section V.
Numerical results are
analyzed in Section VI while the final conclusions are given in Section VII.
\section{The model Hamiltonian. Brief review of frn-RPA }
\label{sec:level2}
Since we are not going to describe realistically some experimental data
but to stress on some specific features of a heterogeneous many 
nucleon system with proton-neutron interaction, we consider a schematic 
Hamiltonian which is very often \cite{Sam,Rad4} used to study the 
single and double beta Fermi transitions: 

\begin{eqnarray}
H & = &
\sum_{jm}{(\varepsilon_{pj}-\lambda_{p})c^{\dag}_{pjm}c_{pjm}}
+\sum_{jm}{(\varepsilon_{nj}-\lambda_{n})c^{\dag}_{njm}c_{njm}}\nonumber\\
&&-\frac{G_{p}}{4}\sum_{jm,j'm'}{c^{\dag}_{pjm}
c^{\dag}_{\widetilde{pjm}}c_{\widetilde{pj'm'}}c_{pj'm'}}
-\frac{G_{n}}{4}\sum_{jm,j'm'}{c^{\dag}_{njm}c^{\dag}_{\widetilde{njm}}
c_{\widetilde{nj'm'}}c_{nj'm'}}\nonumber\\
&&+\chi\sum_{jm,j'm'}{c^{\dag}_{pjm}c_{njm}c^{\dag}_{nj'm'}c_{pj'm'}}
-\chi_{1}\sum_{jm,j'm'}{c^{\dag}_{pjm}c^{\dag}_{\widetilde{njm}}
c_{\widetilde{nj'm'}}c_{pj'm'}}.
\end{eqnarray}
$c^{\dag}_{\tau jm}(c_{\tau jm})$ denotes the creation (annihilation) of a 
$\tau(=p,n)$ nucleon in a spherical shell model state $|\tau;nljm\rangle
=|\tau jm\rangle$ with $\tau$ taking the values $p$ for protons and $n$
for neutrons, respectively. The time reversed state corresponding to 
$|\tau jm\rangle$ is $|\tau{\widetilde{jm}}\rangle=(-)^{j-m}|\tau j-m\rangle$

For what follows it is useful to introduce the quasiparticle ($qp$)
representation, defined by the Bogoliubov-Valatin (BV) transformation:
\begin{eqnarray}
a^{\dag}_{pjm}& = & U_{pj}c^{\dag}_{pjm}-V_{pj}c_{\widetilde{pjm}},\;\;
a_{pjm} =  U_{pj}c_{pjm}-V^{*}_{pj}c^{\dag}_{\widetilde{pjm}}, \nonumber\\
a^{\dag}_{njm}& = & U_{nj}c^{\dag}_{njm}-V_{nj}c_{\widetilde{njm}},\;\;
a_{njm} = U_{nj}c_{njm}-V^{*}_{nj}c^{\dag}_{\widetilde{njm}} .
\end{eqnarray}
which quasi-diagonalizes the first four terms, i.e in the new
representation they are replaced by a set of independent quasiparticles of
energies:
\be 
E_{\tau}=\sqrt{(\epsilon_{\tau}-\lambda_{\tau})^{2}+\Delta_{\tau}^{2}}.
\ee
In the new $qp$ representation, the model Hamiltonian, denoted by $H_q$, 
describes a set of
independent quasiparticles, interacting among themselves through a two body 
interaction determined by the images of the $\chi$ and $\chi_1$ terms
through the BV transformation.

Various many-body approaches have been tested by using not the $qp$ image
of $H$ but another Hamiltonian derived from $H$ by ignoring the
scattering $qp$ terms:
\bea
B^{\dag}(jpn)&=&\sum_{m}a^{\dag}_{pjm}a_{njm}, \nonumber\\
B(jpn)& = &\sum_{m}a^{\dag}_{njm}a_{pjm}.
\eea
and restricting the space of single particle states to a single $j$-state.
Thus, the model Hamiltonian contains, besides the terms for the $qp$ 
independent motion, a two body term which is quadratic in the two
quasiparticle operators $A^{\dag}, A$: 
\be
A^{\dag}(jpn)=\sum_{m}a^{\dag}_{pjm}a^{\dag}_{\widetilde{njm}},
\;\;A(jpn)=(A^{\dag}(jpn))^{\dag}.
\ee

In a previous publication\cite{Rad3}, we showed that going beyond the quasiparticle
random phase
approximation (pnQRPA) through a renormalization procedure, a new degree of
freedom is switched on, which results in having a renormalized pnQRPA boson
operator as a superposition of the operators $A^{\dag}(jpn), A(jpn)$ and 
scattering terms $B^{\dag}(jpn),B(jpn)$. This picture differs from the
standard $pnQRRPA$ approach, where the boson operators involve only the
 operators $A^{\dag}$ and $A$, and is conventionally called as fully
renormalized RPA ($frn-RPA$). Obviously, when the amplitudes of 
scattering terms are dominant, one deals with a new kind of collective $pn$
excitation. 

In order to define clearly the distinct features of the new proton-neutron
($pn$) mode revealed in the present paper a brief description of the
results obtained in a previous publication \cite{Rad3} is necessary.
The equations of motion associated to the many-body Hamiltonian, written in
terms of quasiparticle operators, are determined by the commutators algebra
of the two quasiparticle ($A^{\dag}, A$) and scattering $(B^{\dag},B)$
operators defined by eqs. (2.5) and (2.4) respectively.
Within the $frn-RPA$, the exact commutators are approximated as follows:
\bea
\left[A(jpn),A^{\dag}(jpn)\right ] &=& C^{(1)}_{jpn},
\nonumber \\
\left[B(jpn),B^{\dag}(jpn)\right ] &=& C^{(2)}_{jpn},
\nonumber\\
\left[A(jpn),B^{\dag}(jpn)\right ] &=&\left[A(jpn),B(jpn)\right ] = 0.
\eea

The terms $C^{(1)}_{jpn}, C^{(2)}_{jpn}$ appearing in the r.h.s. of the
above equations are the averages of the corresponding exact commutators, on
the correlated ground state $|0>$:
\be
C^{(1)}_{jpn}=\langle0|1-\hat{N}_{jn}-\hat{N}_{jp}|0\rangle,\: \:
C^{(2)}_{jpn}=\langle0|\hat{N}_{jn}-\hat{N}_{jp}|0\rangle.
\ee
with $\hat{N}_{j\tau}$ standing for the $\tau$ (=p,n) quasiparticle number
operator in the shell j.
The normalized operators
\bea
\bar{A}^{\dag}(jpn)=\frac{1}{\sqrt{C^{(1)}_{jpn}}}A^{\dag}(jpn),\:
\bar{A}(jpn)=\left(\bar{A}^{\dag}(jpn)\right)^{\dag},
\nonumber\\
\bar{B}^{\dag}(jpn)=\frac{1}{\sqrt{|C^{(2)}_{jpn}|}}B^{\dag}(jpn),\:
\bar{B}(jpn)=\left(\bar{B}^{\dag}(jpn)\right)^{\dag},
\eea
satisfy bosonic commutation relations and thereby their equations of
motion are linear:
\be
\left[H_q,\left(\matrix{\bar{A}^{\dag}(jpn) \cr
                        \bar{A}(jpn)        \cr
                         \bar{B}^{\dag}(jpn) \cr
                        \bar{A}(jpn)}\right)\right]
=\sum_{j,j^{\prime}}T^{j,j^{\prime}}\left(\matrix{\bar{A}^{\dag}(jpn) \cr
                        \bar{A}(jpn)        \cr
                         \bar{B}^{\dag}(jpn) \cr
                        \bar{B}(jpn)}\right).
\ee
The matrix $T^{j,j^{\prime}}$ depends on the U and V coefficients as well as
on the strengths $\chi, \chi_1$ of the two body interactions.
The $frn-RPA$ approach defines  a linear combination of the basic operators
$\bar{A}^{\dag}(jpn), \bar{A}(jpn), \bar{B}^{\dag}(jpn), \bar{A}(jpn)$,
\be
\Gamma^{\dag}=\sum_{j}\left[X(j)\bar{A}^{\dag}(jpn)+Z(j)D^{\dag}(jpn)
-Y(j)\bar{A}(jpn)-W(j)D(jpn)\right],
\ee
so that the following commutation relations with its
hermitian conjugate operator and the model Hamiltonian hold:
\bea
\left[\Gamma,\Gamma^{\dag}\right] &=& 1,
\\
 \left[H_q,\Gamma^{\dag}\right] &=& \omega \Gamma^{\dag}.
\eea
The operators $D^{\dag}(jpn)$ are identical with ${\bar{B}}^{\dag}(jpn)$
or $\bar{B}(jpn)$ depending on whether the sign of $C^{(2)}_{jpn}$ is plus or
minus. The equation (2.12) provides a set of homogeneous equations- called
the $frn-RPA$ equations- for the amplitudes $X,Y,Z,W$:
\bea
\left(\matrix{{\cal A} & {\cal B}\cr
              -{\cal B}  &-{\cal A}}\right)\left(\matrix{X\cr Z\cr Y\cr
W}\right) =\omega \left(\matrix{X\cr Z\cr Y\cr W}\right),
\eea
while the equation (2.11) yields the normalization equation
\be
\sum_{j}(X^2(j)+Z^2(j)-Y^2(j)-W^2(j))=1.
\ee
The $frn-RPA$ matrices depend on the renormalization constants $C^{(1)},
C^{(2)}$ which, at their turn, depend on the phonon amplitudes. Therefore,
the equations (2.12) and (2.7) should be self-consistently solved.

In ref. \cite{Rad3} the $frn-RPA$ equations have been solved both for a proton-neutron
dipole-dipole interaction, needed for the description of the
double beta Gamow-Teller decay and for a proton-neutron monopole-monopole interaction
used in the calculation of the rates of the double beta Fermi decay.
Equations obtained in the two cases have some common features which,
for what follows, are worth being enumerated.

\noindent
1) The dimension of the $frn-RPA$ matrix is twice as large as that of the
standard RPA and consequently new solutions show up.

\noindent
2) The solutions characterized by that the largest phonon amplitude is of
type Z define a new class of proton-neutron excitations.

\noindent
3) Due to the attractive character of the two body interaction in the
particle-particle ($pp$) channel, the lowest new state has an energy which
is smaller than the minimal absolute value of the relative energy of the
proton and neutron quasiparticle partner states, related by the operators
$B^{\dag}(jpn), B(jpn)$.

\noindent
4) For the N=Z nuclei, this minimal value is vanishing and therefore the
lowest mode becomes spurious or in other words saying a new symmetry
is open. The new symmetry corresponds to the restriction
$C^{(2)}_{jpn}=0$, i. e. the average of the third component of the 
isospin operator is vanishing.
This means that the system is invariant to rotations around any axes in
the ($X,Y$) plane of the isospin space associated to the (jpn) orbits.

\noindent
5) Important quantitative effects are expected for heavy nuclei having the
proton and neutron Fermi energies lying far apart from each other.

\noindent
6) The presence of the additional states influences also the structure of
the states lying close to those predicted by the standard RPA. Indeed, the
actual normalization condition for the phonon amplitudes implies new
values for the X and Y weights. Consequently, the strengths for $\beta^-$
and $\beta^+$ transitions are shared by the "old"-lying close to the
standard RPA states- and the ``new'' states- for which the amplitudes Z 
are dominant.

\noindent
7) The standard RPA approach is based on the quasi-boson approximation and
therefore it ignores some important dynamic effects (only the terms
$A^{\dag}A^{\dag}, A^{\dag}A, AA$ are considered in an approximative
manner) and moreover
the Pauli principle is violated.
By contrast, within the $frn-RPA$ all the terms of the model Hamiltonian are
taken into account. Also the Pauli principle is, to a certain extent,
restored. Due to this feature, large corrections to the double beta
transition amplitude as well as to the Ikeda sum rule are expected by
changing the RPA to the frn-RPA.

\noindent
8) The equations of motion for the $A^{\dag },A$ and $B^{\dag}, B$
operators are coupled by the terms $A^{\dag}B^{\dag}, A^{\dag}B,
AB^{\dag}, AB$ involved in the quasiparticle Hamiltonian. These terms are
multiplied by the factors $ U_pV_nU_{p^{\prime}}U_{n^{\prime}},
 U_pV_nV_{p^{\prime}}V_{n^{\prime}},  V_pU_nU_{p^{\prime}}U_{n^{\prime}},
 V_pU_nV_{p^{\prime}}V_{n^{\prime}}$ in the $ph-ph$ interaction (the
$\chi$ term)
and by  $U_pU_nU_{p^{\prime}}V_{n^{\prime}},
U_pU_nV_{p^{\prime}}U_{n^{\prime}},  V_pV_nU_{p^{\prime}}V_{n^{\prime}},
V_pV_nV_{p^{\prime}}U_{n^{\prime}}$ in the $pp-hh$ interaction (the $\chi_1$
term), respectively. Note that the coupling terms change the number of
either proton or neutron quasiparticles by two units. The terms
bringing the main contribution to the equations of motion for the
operators $A^{\dag}, A$ commute with $ \hat{N}_{jp}-\hat{N}_{jn}$
but not with $ \hat{N}_{jp}+\hat{N}_{jn}$. By contrary the terms
having the dominant contribution to the equations of motion for the
operators $B^{\dag},B$ commute with $ \hat{N}_{jp}+\hat{N}_{jn}$
and not with $ \hat{N}_{jp}-\hat{N}_{jn}$. None of the two operators,
$ \hat{N}_{jp}-\hat{N}_{jn}$, $ \hat{N}_{jp}+\hat{N}_{jn}$, commutes with
the coupling terms.
Retaining from the $\chi$-interaction the $pp-hh$ terms (those multiplied
by $U_pU_nV_{p^{\prime}}V_{n^{\prime}}$) and from the $\chi_1$ interaction
only the $ph-ph$ terms (those proportional to
$U_pV_nU_{p^{\prime}}V_{n^\prime} $ ) the equations of motion  for the
operators $B^{\dag},B$ are decoupled from those for $A^{\dag}$ and $A$.
One may conclude that the new mode is determined by a combined effect
coming from the $pp-hh$ and $ph-ph$ terms belonging to the $\chi$ and
$\chi_1$ interactions, respectively.

\noindent
9) In the particle representation the $frn-RPA$ phonon operator is a linear
superposition of $ph, hp, pp$ and $hh$ operators.

\noindent
10) In the limit of large $pp$ and negligible $ph$ interactions, the
amplitudes $Z$ can be analytically calculated. The result is that $Z$ is
proportional to either $U_pV_n$ or $V_pU_n$, depending on whether the sign
of $E_p-E_n$ is plus or minus, respectively. When this amplitude prevails
over the other ones, the corresponding mode describes a neutron-hole
proton-particle (or a proton-hole neutron-particle) excitation of the
mother nucleus $(N,Z)$. Therefore the state is associated to the $(N-1,Z+1)$
(or to the $(N+1,Z-1)$) nucleus. In this case the state might be reached by exciting the ground state
through the transition operator $c^{\dag}_pc_n$ (or $c^{\dag}_nc_p$), which is typical for the 
$\beta^-$ ( or $\beta^+$) decay. Since the double beta decay is conceived  as taking place through two
successive $\beta^-$ transitions, one expects that this process is also influenced by considering
this new state as an intermediate state characterizing the odd-odd neighboring nucleus.

\noindent
11) When the $pp$ interaction is small the amplitude Z is proportional to
$U_pU_n$ if $E_p>E_n$ or to $V_pV_n$ in the case $E_p<E_n$.
The new mode characterizes the nucleus $(N+1,Z+1)$ in the first case and
the nucleus ($N-1,Z-1$) in the second situation. The transition operators
 which could excite
these states are obviously of the types $c^{\dag}_pc^{\dag}_n$ and $c_nc_p$, respectively.

Note that the restriction of the phonon operator to the scattering terms
resembles the standard RPA boson operator written in the particle
representation. This comparison has, however, only a formal value since in
the quasiparticle representation there is no Fermi energy and therefore
one cannot speak about quasiparticle-quasihole excitations. Similar
features are met in solid state physics for the description of electron
excitations in narrow energy bands, spin waves and 
plasma oscillations \cite{Hub}. In
nuclear physics, the scattering terms have been also considered but not
for proton-neutron excitations. Indeed, using the thermal response theory,
Tanabe \cite{Tan} studied the charge conserving phonons in nuclear systems
at a finite temperature. It seems that the contribution of the scattering
terms to the charge conserving bosons, does not survive at vanishing 
temperature \cite{Hat}. Moreover, the
dispersion relation for the mode energy cannot be obtained from a
linearized set of equations as it is required by the spirit of the RPA
approach. 

\section{Semi-classical treatment}
\label{sec:level3}

As we already mentioned, the scope of the present paper is to study the
$pn$ mode caused by the quasiparticle scattering terms within a
semi-classical approach. In this formalism the renormalization condition 
(2.6) is missing and therefore the harmonic motion of the new degrees of
freedom hinges on a more physical ground.
Moreover, we address the question whether this mode survives when the
non-scattering terms are switched off. Thus, it is worth to know if such a
mode appears only when the scattering terms accompany the two
quasipatricle operators or it might be determined by the scattering terms
alone. 

From the brief presentation of the previous Section it is clear that the
mode does not appear within the RPA approach. Indeed, it occurred within
the $frn-RPA$ after a consistent renormalization was performed (i. e. not
only the operators $A^{\dag}, A$ where renormalized but also $B^{\dag}$
and $B$). If that mode is a signature of the higher RPA formalisms, then it
should also appear within the semi-classical formalism as well as in the
boson expansion framework. As we shall see the semi-classical approach is
able to predict the mode even in the harmonic approximation, the mode
being associated with the small oscillations of the system around a static
correlated ground state. Moreover, the semi-classical frame is expected to
allow us an intuitive interpretation of this new type of excitation.
 
We recall that the higher order corrections to the standard RPA approach are
frequently studied, with different purposes, using a single j case and
ignoring the scattering terms. The procedure has the advantage that the
resulting Hamiltonian is exactly solvable. Therefore the quality of  
the adopted approximations may be tested by comparing the predictions
with the corresponding exact results. 

 To touch the goal of the present paper
we adopt a similar point of view.
Indeed, if the coupling terms (mentioned at the point 8 of the previous
section) are ignored, the equations of motion for the scattering operators are
decoupled. Moreover the motion of these operators is determined also by an
exactly solvable Hamiltonian, which reads:

\begin{equation}
H^{(q)}_{pn}=E^{'}_{p}{\hat N}_{p}+E^{'}_{n}{\hat N}_{n}+\lambda_{1}
B^{\dag}(pn)B(pn)+\lambda_{2}\big(B^{\dag2}(pn)+B^{2}(pn)\big),
\end{equation}
where the following notations have been used:
\bea
E_p^{\prime}&=&E_p+(\chi_1-\chi)V_p^2,\nonumber\\
E_n^{\prime}&=&E_n+(\chi+\chi_1)V^2_p(V_n^2-U_n^2),\nonumber\\
\lambda_1&=&\chi(U_p^2U_n^2+V_p^2V_n^2)-\chi_1(U_p^2V_n^2+V_p^2U_n^2),
\nonumber\\ 
\lambda_2&=&-(\chi+\chi_1)U_pU_nV_pV_n,\\
{\hat N}_{\tau}&=&\sum_{m}a^{\dag}_{\tau jm}a_{\tau jm}.
\eea
Also, to simplify the notation we omitted the quantum number $j$ for 
the operators $B^{\dag}(jpn), B(jpn)$ as well as for the $U, V$
coefficients and quasiparticle energies.
This model Hamiltonian will be studied within a time dependent variational
formalism. Therefore, some static and dynamic properties will be described
by solving the equations provided by the time dependent variational
principle (TDVP)\footnote{Throughout this paper the units of $\hbar=1$ are used}: 

\begin{equation}
\delta \int_0^t {\langle\Psi|H^{(q)}_{pn}-i\frac{\partial}{\partial t'}|\Psi\rangle}
\,dt' =0.
\end{equation}
If the variational state $|\Psi\rangle$ spans the whole Hilbert space
describing the many-body system, solving the equation (3.4) is equivalent
to solving the time dependent Schroedinger equation, which would be a very
difficult task. In the present paper,
the trial function is taken as:
\begin{equation}
|\Psi\rangle = exp[zB^{\dag}(pn)-z^{*}B(pn)]|NT\; -T\rangle,
\end{equation}
where $|NTT_{3}\rangle$ denotes the common eigenfunction of the quasiparticle 
total number (${\hat N}$), the quasiparticle isospin squared 
(${\hat T}^{2})$, and its 
z-axis projection ($T_{z}$) operators, respectively.
$z$ is a complex function of time and $z^*$, the corresponding complex conjugate function.
We justify this choice by the symmetry properties of the
model Hamiltonian. Indeed, let us note first that $H^{(q)}_{pn}$ commutes 
with the
quasiparticle total number operator. Moreover, it can be written in terms
of the quasiparticle total number operator and generators of the SU(2) isospin algebra

\begin{eqnarray}
\tau_{+1} & = & -\frac{1}{\sqrt 2}B^{\dag}(pn),\nonumber\\
\tau_{-1} & = & \frac{1}{\sqrt 2}B(pn),\nonumber\\
\tau_{0} & = & \frac{1}{2}(\hat N_{p}-\hat N_{n}).
\end{eqnarray}
Due to this property of $H^{(q)}_{pn}$, the function 
$|\Psi\rangle$, which is a coherent state for the SU(2) group, is the
most suitable for a semi-classical treatment.

Before closing this section we would like to write the trial function
in a form which suits better the further purposes. Using the Cambel Hausdorff
factorization\cite{Kir} for the exponential function, as explained in Appendix A,
one obtains:
\begin{eqnarray}
|\Psi\rangle & = & {\cal N} e^{\alpha B^{\dag}(pn)}|NT\;-T\rangle,
\nonumber\\
{\cal N}& = &(1+\alpha^{*}\alpha)^{-T}. 
\end{eqnarray}
where $\alpha$ depends on the polar coordinates, $z=\rho e^{i\varphi}$:
\be
\alpha=\tan(\rho)e^{i\varphi}.
\ee
\section{Equations of motion}
\label{sec:level4}
In order to write the equations of motion provided by the TDVP (3.4),
we need the matrix element of $H^{(q)}_{pn}$ as well as of the time derivative operator, 
$\frac{\partial}{\partial t}$. These can be evaluated by direct calculation, using the
expressions (3.5) when the average of $H^{(q)}_{pn}$ is considered and (3.7) for the
classical action. The result is:
\begin{eqnarray}
\langle\Psi|H^{(q)}_{pn}|\Psi\rangle& =& -T(E_{p}^{\prime}-E_{n}^{\prime}+2\lambda_{1})
+\frac{N}{2}(E_{p}^{\prime}+E_{n}^{\prime})+2T(E_{p}^{\prime}-E_{n}^{\prime}+2\lambda_{1})
\frac{\alpha^{*}\alpha}{1+\alpha^{*}\alpha}\nonumber\\
&&+2T(2T-1)\left[\lambda_{1}\frac{\alpha^{*}\alpha}
{(1+\alpha^{*}\alpha)^{2}}+\lambda_{2}\frac{\alpha^{*2}+
\alpha^{2}}{(1+\alpha^{*}\alpha)^{2}}\right],
\nonumber\\
\langle\Psi|\frac{\partial}{\partial t}|\Psi \rangle &=&
T\frac{\alpha^{*}\stackrel{\bullet}{\alpha}-\stackrel{\bullet}{\alpha}^{*}
\alpha}{1+\alpha^{*}\alpha}.
\end{eqnarray}
Considering $\alpha,\alpha^*$ as classical phase space coordinates, the
TDVP equation (3.4) yields the following classical equations of motion, 
describing the nuclear system:

\begin{eqnarray}
\frac{\partial{\cal H}}{\partial\alpha}&=&-2i
\frac{T\stackrel{\bullet}{\alpha}^{*}}{(1+\alpha^{*}\alpha)^{2}}, \nonumber\\
\frac{\partial{\cal H}}{\partial \alpha^{*}}&=&2i
\frac{T\stackrel{\bullet}{\alpha}}{(1+\alpha^{*}\alpha)^{2}}.
\end{eqnarray}
Here ${\cal H}$ denotes the classical energy function:
\be
{\cal H}=\langle\Psi|H^{(q)}_{pn}|\Psi\rangle.
\ee
In order to quantize the classical trajectories satisfying the equations
(4.2) as well as to have an one to one correspondence between the classical
and quantal behaviors of the nucleon system, it is convenient to chose
those conjugate variables which bring the equations of motion in a
canonical Hamilton form. A possible choice of the coordinates with the
above mentioned property is
\begin{eqnarray}
r&=&\frac{2T}{1+\alpha^{*}\alpha},\\
\psi&=&-\frac{1}{2i}(\ln \alpha-\ln \alpha^{*})=-\varphi.
\end{eqnarray}
Indeed, in the new variables the classical equations read:
\begin{eqnarray}
\frac{\partial{\cal H}}{\partial r}&=&-\stackrel{\bullet}{\psi},\nonumber\\
\frac{\partial{\cal H}}{\partial \psi}&=&\stackrel{\bullet}{r}.
\end{eqnarray}
with the classical energy:
\bea
{\cal H}&=&T(E_{p}^{\prime}-E_{n}^{\prime}+2\lambda_{1})+\frac{N}{2}
(E_{p}^{\prime}+E_{n}^{\prime})
\nonumber\\
&&-(E_{p}^{\prime}-E_{n}^{\prime}+2\lambda_{1})r+\frac{2T-1}{2T}r(2T-r)
(\lambda_{1}+2\lambda_{2}\cos2\psi).
\eea
Note that $r$ has the significance of a generalized coordinate while $\psi$
that of generalized linear momentum.
Due to the generalized momentum $\psi $, the equations motion are not
linear and therefore analytical solutions are not obtainable.
The equations can however be approximatively solved if they are linearized around the 
minimum point of the energy function:
\be
\stackrel{\circ}{r}=T\left[1-\frac{E_{p}^{\prime}-E_{n}^{\prime}+
2\lambda_{1}}{(2T-1)(\lambda_{1}-2\lambda_{2})}\right],
\;\;
\stackrel{\circ}{\psi}=\frac{\pi}{2}. 
\ee
In order that the minimum exits, it is necessary that the generalized 
coordinates satisfy a
consistency condition, required by the definition range of $r$:
\be
0\leq\stackrel{\circ}{r}\leq2T.
\ee
By means of (4.8), this provides a constraint for the strengths of the two
body interactions.
The linearized equations, written in terms of the deviations
\be
q=r-\stackrel{\circ}{r},\;  p=\psi-\stackrel{\circ}{\psi},
\ee
are of harmonic type:

\begin{eqnarray}
-\stackrel{\bullet}{p}=-2\frac{2T-1}{2T}(\lambda_{1}-2\lambda_{2})q,
\nonumber\\
\stackrel{\bullet}{q}=4\frac{2T-1}{T}\stackrel{\circ}{r}
(2T-\stackrel{\circ}{r})\lambda_{2}p.
\end{eqnarray}

These describe a harmonic motion for the conjugate coordinates, with the
angular frequency:

\begin{equation}
\omega=2\frac{2T-1}{T}[-\lambda_{2}(\lambda_{1}-2\lambda_{2})\stackrel{\circ}
{r}(2T-\stackrel{\circ}{r})]^{\frac{1}{2}}.
\end{equation}
The condition that $\omega$ is a real quantity brings an additional constraint for
the strength parameters $\chi,\chi_1$:

\be
\chi\geq\chi_1\left(\frac{U_pV_n-V_pU_n}{U_pU_n+V_pV_n}\right)^2.
\ee
As we said already before, the schematic model has the advantage, over the
realistic formalisms, that allows us to compare the approximative solutions
with the exact one. For the particular Hamiltonian used in the present paper,
the exact eigenvalues can be obtained by diagonalization in the basis
{$|NTM\rangle$}. Indeed, in this basis the model Hamiltonian has the
following non-vanishing matrix elements:
\bea
\langle NTM|H^{(q)}_{pn}|NTM\rangle &=&\frac{1}{2}(E_p^{\prime}+E_n^{\prime})N
+(E_p^{\prime}-E_n^{\prime})M+
\lambda_1(T+M)(T-M+1),\nonumber\\
\langle NTM+2|H^{(q)}_{pn}|NTM\rangle &=&\lambda_2\left[(T-M-1)(T-M)(T+M+1)(T+M+2)
\right]^{\frac{1}{2}},
\nonumber\\
\langle NTM|H^{(q)}_{pn}|NTM+2\rangle &=&\langle NTM+2|H^{(q)}_{pn}|NTM\rangle.
\eea

\section{The renormalized RPA and boson expansion}
\label{sec:level5}

Within the RPA approach, the renormalization of the quasiparticle mean
field due to the two quasiparticle interactions is usually ignored. 
Therefore the Hamiltonian considered is:

\begin{equation}
H_{qp}=E_{p}{\hat N}_{p}+E_{n}{\hat N}_{n}+\lambda_{1}
B^{\dag}(pn)B(pn)+\lambda_{2}\big(B^{\dag2}(pn)+B^{2}(pn)\big),
\end{equation}
The operators $B^{\dag},B$ satisfy the commutation relation:
\be
[B(pn),B^{\dag}(pn)]={\hat N}_n-{\hat N}_p.
\ee
If the r.h. side of the above equation is replaced by its
average on the ground state,
\be
C=\langle  0|{\hat N}_n-{\hat N}_p|0\rangle
\ee 
which is to be determined, then the operators
$B,B^{\dag}$ become bosons, after the following renormalization
\be
{\widetilde {B}}^{\dag}(pn)=\frac{1}{\sqrt{C}}B^{\dag}(pn),\;
{\widetilde {B}}(pn)=\frac{1}{\sqrt{C}}B(pn),
\ee
if $C$ is positive, while for negative $C$ the renormalized operators are:
\be
{\widetilde {B}}^{\dag}(pn)=\frac{1}{\sqrt{|C|}}B(pn),\;
{\widetilde {B}}(pn)=\frac{1}{\sqrt{|C|}}B^{\dag}(pn).
\ee
Suppose, for the time being, that $C>0$. If that is not the case
the corresponding calculations can be worked out in a similar way.
The equations of motion for the renormalized operators are:
\bea
\left[H_{qp},{\widetilde {B}}^{\dag}(pn)\right]& = & (E_p-E_n+\lambda_1C)
{\widetilde{B}}^{\dag}(pn) +2\lambda_2C{\widetilde{B}}(pn),
\nonumber \\
\left[ H_{qp},{\widetilde{B}}(pn)\right]& =& -2\lambda_2C{\widetilde{B}}^{\dag}(pn)
-(E_p-E_n+\lambda_1C){\widetilde{B}}(pn) .
\eea
Since the equations are linear in ${\widetilde {B}}^{\dag}(pn)$ and
${\widetilde{B}}(pn)$, one can define the phonon operator
\be
\Gamma^{\dag}=X\widetilde {B}^{\dag}(pn)-Y\widetilde {B}(pn),
\ee
with the amplitudes determined such that the following equations are
fulfilled: 
\bea
\left[H_{qp},\Gamma^{\dag}\right]&=&\omega\Gamma^{\dag},\nonumber\\
\left[\Gamma,\Gamma^{\dag}\right]&=&1.
\eea
The first equation provides the dispersion equation for the mode energy
\be
\omega=\left[(E_p-E_n+\lambda_1C)^2-4\lambda_2^2C^2\right]^{\frac{1}{2}},
\ee
while the second one the normalization relation for phonon amplitudes:
\be
X^2-Y^2=1.
\ee
The renormalized RPA vacuum is defined by
\be
\Gamma|0\rangle=0.
\ee
The solution of the above equation is:
\be
|0\rangle=e^{-\frac{1}{8}(\frac{Y}{X})^2}e^{\frac{Y}{2X}{\widetilde
{B}}^2} |NT,-T\rangle.
\ee
Then the renormalization constant $C$ can be exactly evaluated:
\be
C=2T-2+\frac{2}{X^2}.
\ee
Since $T\ge1$, the constant C is always positive.
The equations of motion allow us to express the amplitude Y in terms of X:
\be
Y=\frac{1}{2\lambda C}[\omega-(E_p-E_n+\lambda_1C)]X,
\ee
which together with the normalization condition (5.10) determines fully the
amplitudes X and Y in terms of C and $\omega$. Inserting the result for $X$
into the equation (5.13),
one obtains an equation for C as a function of $\omega$.
This and eq.(5.9) form a set of two nonlinear equations for the unknowns
$\omega$ and $C$.

As we mentioned before, another way to improve the RPA treatment is to use
the boson expansion concept. Through this procedure, the $SU(2)$ 
algebra, with the fermionic generators $\tau_{\pm 1}, \tau_0$ defined by
eq.(3.6), is mapped to
a boson $SU(2)$ algebra, generated by $\hat {T}_{\pm 1}, \hat {T_0}$.
Denoting by $b^+,b$ a pair of boson operators, the
$SU(2)$ algebra generators $\hat {T}_{\pm 1},\hat {T}_0$ can be constructed
as function of $b^+$ and b. The resulting expressions are conventionally
called as the boson expansion of the fermionic generators, respectively.
There are three distinct boson mappings for the fermionic SU(2) algebra found
by Holstein-Primakoff \cite{Hol}, Dyson \cite{Dy} and one of the present
authors (A. A. R.)\cite{Rad5}, respectively.
For the present purpose here we use the Holstein-Primakoff (HP) expansion:
\bea
\hat{T}_{+1}&=&-\sqrt{T}b^+\left(1-\frac{b^+b}{2T}\right)^{\frac{1}{2}},   
\nonumber\\
\hat{T}_{-1}&=&\sqrt{T}\left(1-\frac{b^+b}{2T}\right)^{\frac{1}{2}}b,
\nonumber\\
\hat{T}_0&=&b^+b-T.
\eea
By a direct calculation it can be checked that, by this mapping, to the
operator $\tau ^2$ it corresponds a C-number:
\be
\hat{T}^2=T(T+1).
\ee
The fermion Hamiltonian $H_{qp}$ commutes with the quasiparticle
total number and the same is true for the generators $\tau_{\pm 1},\tau_0$.
Therefore the image of the quasiparticle total number operator through
the  HP mapping is invariant against any rotation in the isospin space and
consequently, according to the above equation, is a C-number.
Apart from an additive constant, the image of $H_{qp}$ through the HP
boson expansion is:
\be
H^{(b)}_{qp}=(E_p-E_n)\hat{T}_0-2\lambda_1\hat{T}_{+1}\hat {T}_{-1}
+2\lambda_2(\hat {T}^2_{+1}+\hat {T}^2_{-1}).
\ee
Making use of eqs. (5.15), the boson mapping of $H_{qp}$ is a infinite
series in the bosons $b^+, b$, due to the square root operators.
Expanding the square root operators and truncating the result at the
second order in bosons, the boson Hamiltonian becomes:
\be
H^{(b)}_{qp;2}=(E_p-E_n+2\lambda_1T)b^+b+2\lambda_2T({b^+}^2+b^2).
\ee
For a limited range of the interaction strength, this Hamiltonian can be
diagonalized through a canonical transformation:
\bea
b^+&=&UB^++VB,
\nonumber\\ 
b&=&UB+VB^+,
\nonumber\\
1&=&U^2-V^2.
\eea
The restriction that the "dangerous" terms have a vanishing strength
yields the expression for the transformation coefficients and the
coefficient, $\omega_1$, of the diagonal term $B^+B$:
\bea
\left(\matrix{U\cr V}\right)&=&\frac{1}{\sqrt{2}}\left[\mp
1+\frac{|E_p-E_n|}{\sqrt{(E_p-E_n+2\lambda_1T)^2 -16\lambda_2^2T^2}}
\right]^{\frac{1}{2}},
\nonumber\\
\omega_1&=&\left[(E_p-E_n+2\lambda_1T)^2
-16\lambda_2^2T^2\right]^{\frac{1}{2}}. 
\eea
Comparing the expressions of $\omega_1$ (5.20) and $\omega$ (5.9),
one sees that the two energies are identical for the limiting case of
$X=1$, which is met when $\lambda_2=0$ (see eqs. (5.9) and (5.14)).

At this stage it is worthwhile to make the following remarks:
a) When the HP boson expansion of the model Hamiltonian is truncated at
the second
order terms in bosons, the quasiparticle total number operator is no
longer a C number. Therefore the contribution of this term should have
been considered in a consistent manner.
Moreover the truncation is justified only for large values of the total
isospin T.
 b) The same inconsistency appears
in the calculation of the renormalization constant C. Indeed the
expression (5.14) is exact and therefore includes all contributions coming
from the infinite boson series of the correlated ground state, given by
(5.14). 
c) Since the boson mapping (5.15) is an unitary transformation, the
exact eigenvalues of $H_{qp}$ are reproduced by diagonalizing the boson
expanded Hamiltonian $H^{(b)}_{qp}$.
For the second order truncated Hamiltonian, the canonical transformation 
breaks down at a critical value of the attractive interaction strength.
However the diagonalization procedure is able to find the eigenvalues for
any strength of the attractive interaction. The resulting energies exhibits a phase transition
(the first derivative has a jump) at the critical value of the strength.
If the second branch of the energy curve could also be approximated by an
harmonic mode, describing small oscillations of the classical system around
a stationary state, this is still an open question \cite{Rad4}. 
d) The HP boson representation provides  for the harmonic
mode the interpretation of an wobbling motion of the system around the 
total isospin. e) The HP boson expansion is justified (in the sense that  
some eigenvalues of the truncated Hamiltonian are close to the
corresponding exact ones) when the rotation axis in the isospin space is
close to the quantization axis (z axis), which is usually taken as the axis
to which the maximum ``moment of inertia'' corresponds. If the angle between the 
rotation axis and z-axis is large the harmonic energy may collapse.
In this case the quantization axis should be chosen as one of the X and Y axes
depending on the magnitude of the strength of the $\tau_x^2$ and
$\tau_y^2$ terms from the quasiparticle Hamiltonian $H^{(q)}_{pn}$.
In this case the boson representation suitable for the low order
description should be of Dyson type \cite{Rad6}. The harmonic approximation 
for the new representation describes also a wobbling motion
of a frequency equal to the square root of the product of the inverse of the non-maximal 
moments of inertia normalized to the inverse of the maximal moment of inertia.

\section{Numerical results}
\label{sec:level6}
The formalism described in the previous sections, has been applied to the
case $j=\frac{19}{2}$. On the proton level, 6 protons are distributed 
while in the neutron level, 14 neutrons. Alike nucleons interact with each
other through pairing forces whose strength are $G_p=0.2 {\rm MeV}$ and $G_n=0.4$
MeV. From the pairing equations it results the following expression for
the quasiparticle energy:
\be
E_{\tau}=\frac{1}{2}G_{\tau}\Omega,\; \Omega=\frac{2j+1}{2}.
\ee
With the data specified above the result for the quasiparticle energy is:
\be
E_p=2\; {\rm MeV},\;\;E_n=1\;\;{\rm MeV}.
\ee
According to our previous study, the renormalized RPA ground state
involves a small number of quasiparticles. For example, for a small
strength of the particle-particle interaction, the quasiparticle
total number is about 2 while for large values of the above mentioned
strength the number may reach the value 4. Due to this behavior of the
correlated ground state we considered for the isospin carried by the 
quasiparticles in the ground state, alternatively the values 1 and 2.
Although these values vary with increasing the particle particle strength
we kept them constant.

The numerical analysis refers to the dependence of the energy $\omega$
of the new nuclear mode, on the strengths of the $ph$ and $pp$ monopole
interactions, $\chi, \chi_1$. Aiming at showing how good is the
semi-classical approach for this new type of pn excitation, we calculated 
also the
exact eigenvalues of the model Hamiltonian, by diagonalizing the associated
matrix (4.14) within the basis $|NTM\rangle$.
The results are shown in Fig. 1 and Fig. 2.
From Fig. 1, one notices that the harmonic mode collapses for a critical
value of the attractive interaction $\chi_1$.
This critical value is certainly depending on the repulsive interaction 
strength. The larger is that strength the larger the critical value.
In Fig. 1, we have also plotted the normalized energy for the first excited
state. There are intervals for $\chi_1$ where the energy of the harmonic
mode approximates reasonably well the exact excitation energy. Moreover, for
two values of the strength parameter, the exact solutions are precisely 
reproduced. For the case T=2, the two energies, exact and $\omega$, are
the same for $\chi_1=0$ and $\chi_1=0.3$ for $\chi=1$  and $\chi=0.5$
respectively, but the curves are going apart for the first part of interval
and then converge to an intersection point close to the critical value.  

The peculiar feature of $\omega$ as a function of
$\chi_1$, which distinguishes it from the standard RPA modes,
consists of its non-monotonic behavior with respect to the increase of
the strength of the $pp$ interaction. The reason is that in the common
cases the mean field is constant when the two body interaction is varied,
while here by changing $\chi_1$ we change also the minimum point for energy
and therefore another mean field is obtained. It is interesting to notice
that although the ph interaction,  the $\chi$ term, is kept constant,
the change of the mean field is equivalent to an increase of the effective
$ph$ interaction until $\omega$ reaches the maximum value from where the
attractive component of the two body interaction prevails.

In Fig. 2, the energies $\omega$ and the normalized
energy of the first excited state are shown as function of $\chi$, the
strength parameter of the $ph$ interaction. Both energies are
monotonically increasing with the increase of the interaction strength. 
In contrast to what happens in the case of $\chi_1$ dependence, here
the change of the mean field by changing the energy minimum does not change
the repulsive character of the $\chi$ interaction. The agreement between
$\omega$ and the exact energy of the first excited state is reasonable good.

In Fig. 3. the energies characterizing the harmonic mode predicted by the
renormalized RPA and semi-classical method are plotted as function of 
$\chi_1$. Also, the exact energy of the first excited state is presented.
Although they have different trends, the semi-classical and renormalized
RPA energies are not far from each other for $\chi<0.45$. At the 
critical value $\chi$=0.57 the energy yielded by the
renormalized RPA is going very fast to zero. This behavior is specific
to the present model where only the scattering terms are considered.
Indeed, if the phonon operator includes both the two quasiparticle and
scattering terms, the corresponding mode collapses for larger $\chi_1$.
In the semi-classical treatment this happens only for very large $\chi_1$ 
since the static ground state is changed by increasing $\chi_1$. 
The result obtained with the truncated HP boson expanded Hamiltonian
(see eq. (5.20)) is very close to the result shown in Fig. 3 for the 
renormalized RPA procedure.

Comparing the results from Fig. 1a and Fig. 3, we remark on 
the following features.
While the renormalized RPA energy collapses at a relatively small value of
$\chi_1$, the mode energy predicted by the semi-classical formalism
vanishes for a very large $\chi_1$, far beyond the realistic value, which is
$\chi_1=\chi$. This feature is a consequence of
changing the static ground state with $\chi_1$.
The energy behavior provided by the semi-classical method is also
different from that predicted by the standard renormalized pnQRPA
(see for example ref. 9) where the mode energy is a monotonic function of
$\chi_1$ and goes asymptotically to zero.

In this context we recall that the $frn-RPA$ breaks down \cite{Rad3}
before the standard RPA does, and that happened due to the fact that the
lowest $frn-RPA$ energy is that associated with the new collective mode.
From the present calculations one sees that {\it this is not true within the 
semi-classical approach and therefore including the scattering terms in
the expression of the phonon operator does not prevent the treatment of
the many-body system
for a realistic value of the $pp$-interaction strength.}

The vanishing energies for the new mode, shown in Figs. 1a and 3 suggest
that a phase transition occurs according to the corresponding formalisms. 
As we already mentioned this is clearly revealed if one diagonalizes the
Hamiltonian given be eq. (5.18) \cite{Sam,Rad4}. In the renormalized RPA
procedure the new phase is determined by a new minimum of the classical
energy associated to $H^{(b)}_{pn;2}$, reflecting the fact that the
$\lambda_2$ term is the dominant one for these values of $\chi_1$.
In the full-line and dotted-line curves of Fig. 1a, the corresponding energies
also vanish at certain critical values which result in having again a phase transition.
This is reflected in the
curve obtained by exact calculations, by the fact that the energy is minimum for the critical strength.
The increasing branch shown by the exact calculations (corresponding to
the second nuclear phase) might be semi-classically described by changing
the trial function, involved in the time dependent variational equations,
by rotating it (in the isospin space) with an angle which corresponds to the 
orientation of the axis of maximum ``moment of inertia''.

It is remarkable that the far intersection points of the curves obtained
by semi-classical and exact calculations respectively, are lying close to
the critical values of the semi-classical description. Also, the first 
intersection point is not far from the critical value of the
renormalized RPA treatment. In the classical treatment this feature is
well known \cite{Ghe}. Indeed, in the above quoted reference it is shown,
for a triaxial rotor cranked on an arbitrarily oriented axis, that for certain 
critical values of the strength parameters, the period of the harmonic orbits 
is equal to the period characterizing the motion on the closed exact orbit.

\section{Conclusions}
\label{sec:level7}
The main result of this paper refers to the existence of an harmonic mode
determined by the scattering quasiparticle terms, which are usually
neglected in the standard RPA approach. 

The new mode is described within a time dependent variational formalism with an exactly solvable 
many-body Hamiltonian. The variational state is a coherent state for the underlying symmetry 
group, which is the SU(2) group.
A pair of classical canonical conjugate coordinates, which bring the equations of 
motion to the Hamilton form, is found. The classical energy has an interesting structure.
It is quadratic in coordinate but highly non-linear in the conjugate momentum.
Therefore one finds first the stationary point which minimizes the energy, and then linearizes the
equations of motion around the minimum point in the classical phase space. The solution for the linearized equations is
harmonic and its time period determines the energy of the new mode. Despite the fact the classical system has an harmonic
motion, the mode does not exist in the standard RPA approach. In this sense one may say that the present description corresponds to a
''renormalized RPA''. However as we have seen, by comparing the corresponding predictions, the renormalization involved in 
the semi-classical description is completely different from the renormalization described in Section V as well 
as from the boson expansion method.

It is known the fact that the topological structure of the energy surface depends on the strength parameters involved in the model Hamiltonian.
Thus, in the parameters space one can define several regions, each of them corresponding to a distinct nuclear phase.
Having this in mind, we studied  the behavior of the new mode energy when the strength parameter for the $pp$
interaction ($\chi_1$) is varied. A particular feature for the semi-classical description is that the energy is not monotonic 
decreasing function of $\chi_1$, but it increases in the first part of the interval, reflecting that here the $ph$ two body interaction prevails,
reaches a maximum value, then decreases and finally vanishes. This property is caused by that for each $\chi_1$
a new ground state is determined. This aspect is missing in both the renormalized RPA and boson expansion procedures.

Since the model Hamiltonian resembles the triaxial rotor which was semi-classically studied by one of the present authors (A. A. R)
in refs. \cite{Rad6,Ghe}, the interpretation of the new mode is imported from there. Thus, the new mode describes a wobbling motion around
a given total isospin.

The vanishing energy is a sign for a phase transition. In the first phase the rotation axis, in the isospin space 
lies close to the z-axis, which has the maximum moment of inertia in the region of small $\chi_1$, while for $\chi_1$ 
larger than the critical value (where the energy vanishes), the rotation axis lies closer to the (X,Y) plane in the isospin space.
While the first phase may be described by a HP boson expansion formalism, for the second phase the Dyson boson representation is the proper one [20].
In the semi-classical approach, the new phase might be described by changing the trial function associated to the first phase,
through a rotation which brings the z-axis to the actual axis of maximal ``moment of inertia''.

The occurrence of the phase transition can be noticed also in the curve showing the exact first excitation energy as function of $\chi_1$.
Indeed at the critical value of $\chi_1$, this curve exhibits a minimum.

Another critical values of $\chi_1$ are those where the mode energy is equal to the exact excitation energy produced by the diagonalization procedure.
For these values the linearization does not affect at all the period of the exact closed classical orbit. As a matter of fact, for $\chi_1$ lying close to 
these points the linearization are best justified. It is interesting to notice that these values of $\chi_1$
 lie however close to the values where the phase transitions in the
semi-classical treatment (the far intersection point) and the renormalized RPA approach (the near intersection point) take place. 
This observation allows us to conclude that
the semi-classical approach works very well for the values of $\chi_1$ where the renormalized RPA breaks down and that the interval where
the linearization procedure does not work, ending with the critical value where the semi-classical energy vanishes, is very narrow.

The energy of the new mode vanishes for a value of $\chi_1$ which is far beyond the physical value ($\chi_1=\chi$). 
In this way the drawback of the $frn-RPA$, of breaking down earlier than the standard RPA does, is removed.

How could  the new state be populated? We identified the transition operators which could excite the new state from the ground state.
The conclusion is that these state can be seen either in a $\beta^-$ (or $\beta^+$) decay or in a deuteron transfer reaction experiment. 

The coupling of this mode to other collective states will be studied in
a subsequent paper using a realistic interaction and a large model space
for the single particle motion.

\vskip0.5cm
{\bf Acknowledgement}
\vskip0.5cm
\noindent
One of us (B. C.) wants to thank Prof. Amand Faessler for hospitality
in Institute of Theoretical Physics of Tuebingen University where 
a part of this work was performed. A. A. R. thanks Prof. Faessler for
reading the manuscript and valuable remarks concerning the structure of
the variational function.
 
\section{Appendix A}
\label{sec:level8}
Here to derive the factorization of the trial function
$|\Psi\rangle$. To this purpose we address the following more general
question. Which are the t-functions $A(t), B(t), C(t)$ satisfying the
equation
\be
e^{t[(z\sum_{m}a_{pm}^{\dag}a_{nm}-z^{*}\sum_{m}a_{nm}^{\dag}a_{pm})]}
=e^{A(t)a_{p}^{\dag}a_{n}}e^{C(t)(\hat N_{p}
-\hat N_{n})}e^{B(t)a_{n}^{\dag}a_{p}},
\ee
with the initial conditions
\begin{equation}
A(0)=B(0)=C(0)=0
\end{equation}
and t a real parameter. Once we solve this problem
the needed factorization is obtain from (4.1) for $t=1$.
Taking the first derivative of the eq.(6.1), with respect to t,  and
identifying the coefficients of the similar operators one obtains the
following system of differential equations for the three unknown functions,
$A(t), B(t), C(t)$ :

\begin{eqnarray}  
\stackrel{\bullet}{z} &=& \stackrel{\bullet}{A}-2A
\stackrel{\bullet}{C}-\stackrel{\bullet}{B}A^{2}e^{-2C(t)},\nonumber\\
0&=& \stackrel{\bullet}{C}+\stackrel{\bullet}{B}Ae^{-2C(t)},\nonumber\\
-\stackrel{\bullet}{z}^* &= &\stackrel{\bullet}{B}e^{-2C(t)}. 
\end{eqnarray}
Eliminating the functions $B, C$ from these equations, one obtains the following equation for 
$A(t)$.
\begin{equation}
\stackrel{\bullet}{z}=\stackrel{\bullet}{A}-A^{2}\stackrel{\bullet}{z}^*
\end{equation}
which admits the solution:
\be
A(t)=\tan(\rho t)e^{i\varphi}.
\ee
Here the polar coordinates $(\rho,\varphi)$($z=\rho e^{i\varphi}$) have been used.
Inserting the result for $A(t)$ in the eq. (6.3), the equations for the remaining
functions can be easily integrated. The result is:
\bea
C(t)&=&-\ln \big( \cos(\rho t) \big)\nonumber\\
 B(t)&=&\tan(\rho t) e^{-i\varphi}.
\eea

For the sake of simplifying the writing, hereafter the following
notation will be used:
\be
\alpha=A(1)
\ee
Using these results the trial function can be written as:

\bea
|\Psi\rangle & = & e^{-2C(1)T}e^{A(1)B^{\dag}(pn)}|NT\;-T
\rangle\equiv {\cal N} e^{A(1)B^{\dag}(pn)}|NT\;-T\rangle
\eea
where $\cal N$ denotes the normalization factor:
\begin{equation}
{\cal N} = e^{-2C(1)T}=e^{2 \ln(\cos \rho)T}={(1+|\alpha|^2)}^{-T}.
\end{equation}
\nopagebreak

\vspace*{-2cm}
\begin{figure}
\centerline{\epsfig{figure=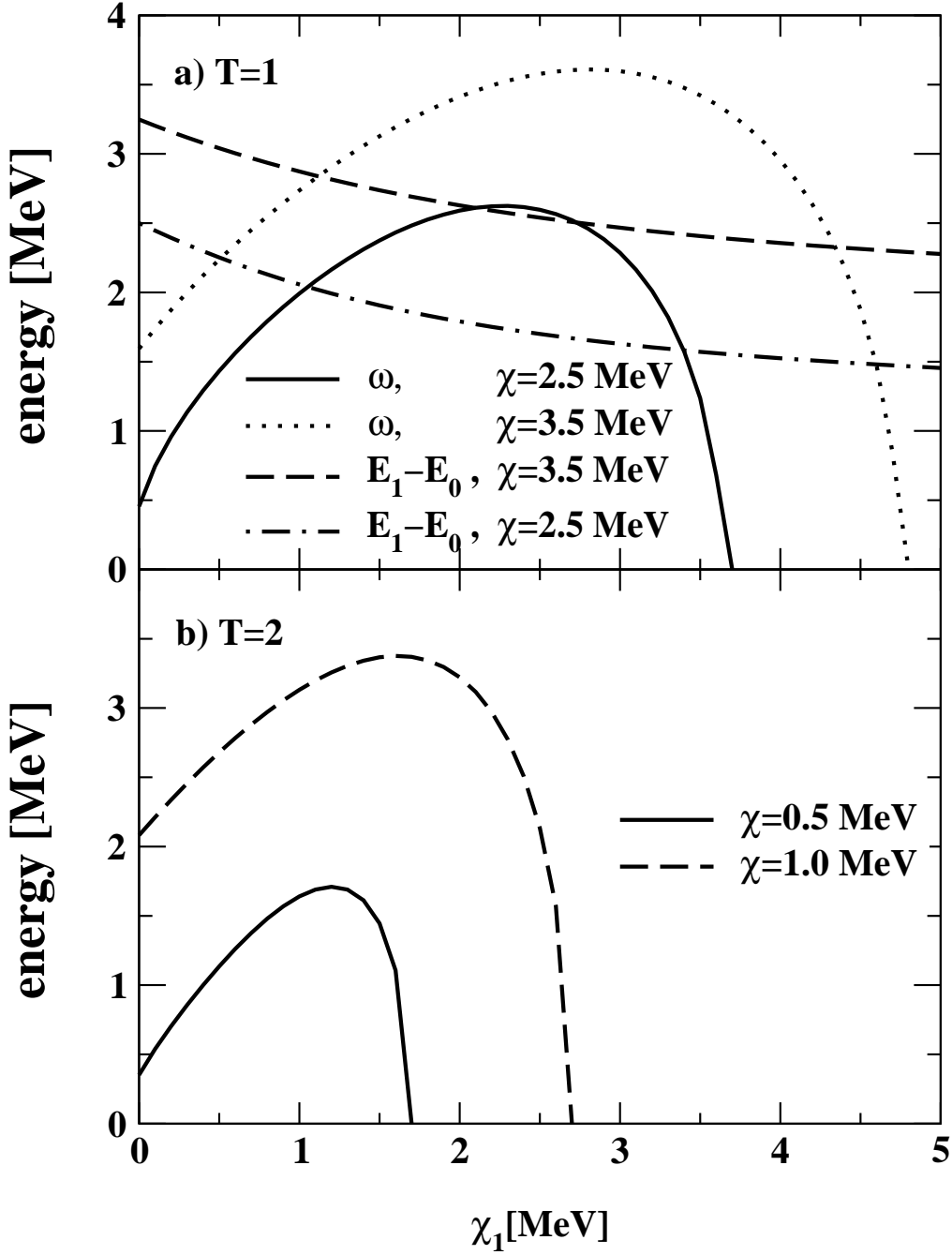,width=14cm,bbllx=2.5cm%
bblly=2.5cm,bburx=18cm,bbury=24.5cm,angle=0}}
\vspace{-2cm}
\caption{The energy of the harmonic mode given by the eq. (4.12) and
the energy of the first excited state normalized to the ground state
are plotted as function of $\chi_1$ for $T=1$ (a)) and $T=2$ (b)) and
several values of the particle-hole interaction strength, $\chi$.}
\label{Fig 1.}
\end{figure}
\vspace*{-2cm}
\begin{figure}

\centerline{\epsfig{file=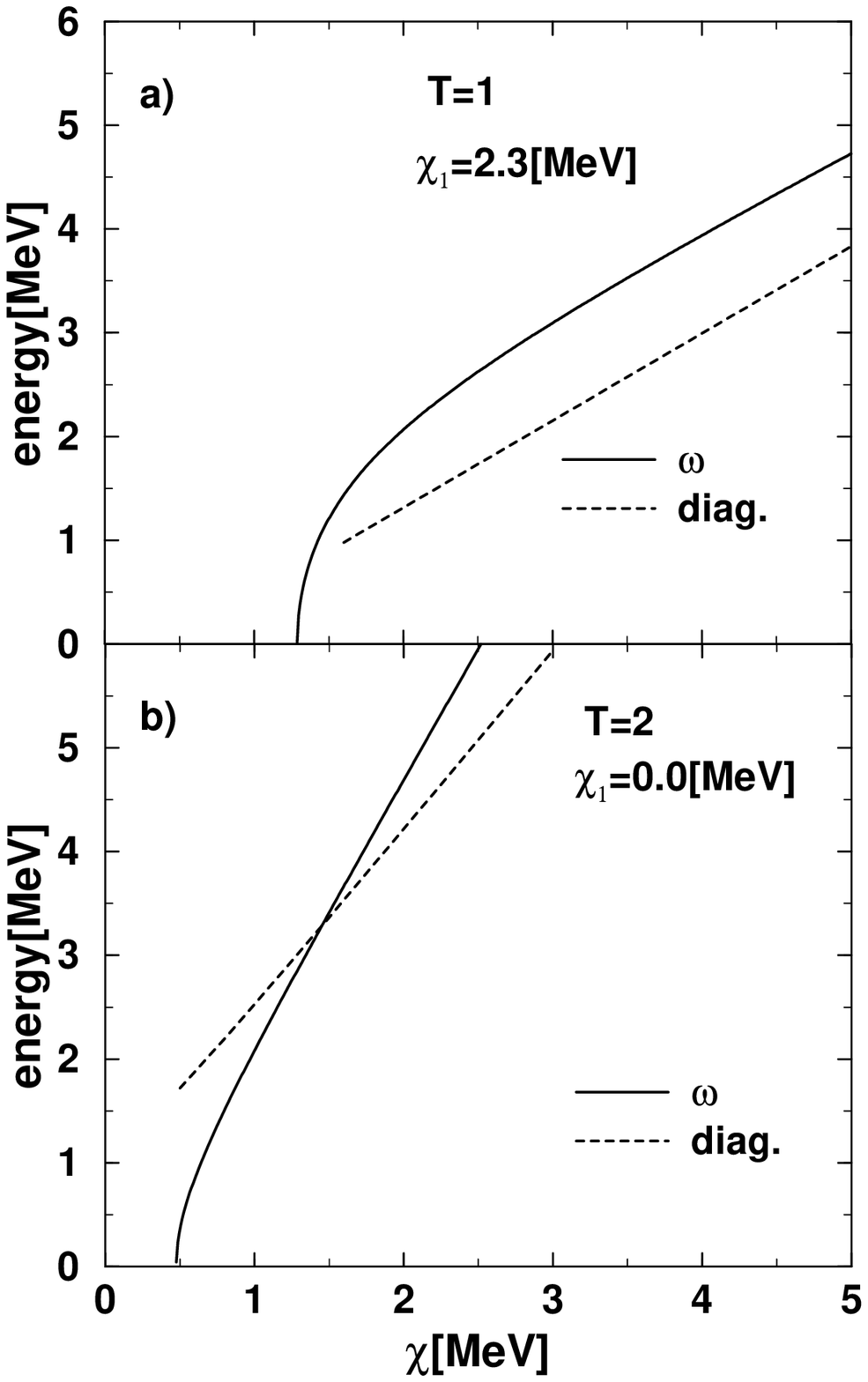,width=16cm}}
\vspace*{-2cm}
\caption{The energy of the harmonic mode given by the eq. (4.12) and
the energy of the first excited state normalized to the ground state
are plotted as function of $\chi_1$ for $T=1, \chi_1=2.3 MeV$ (a)) 
and $T=2, \chi_1=0.0 MeV$ (b)). 
}
\label{Fig 2.}
\end{figure}

\vspace*{2cm}
\begin{figure}

\centerline{\psfig{figure=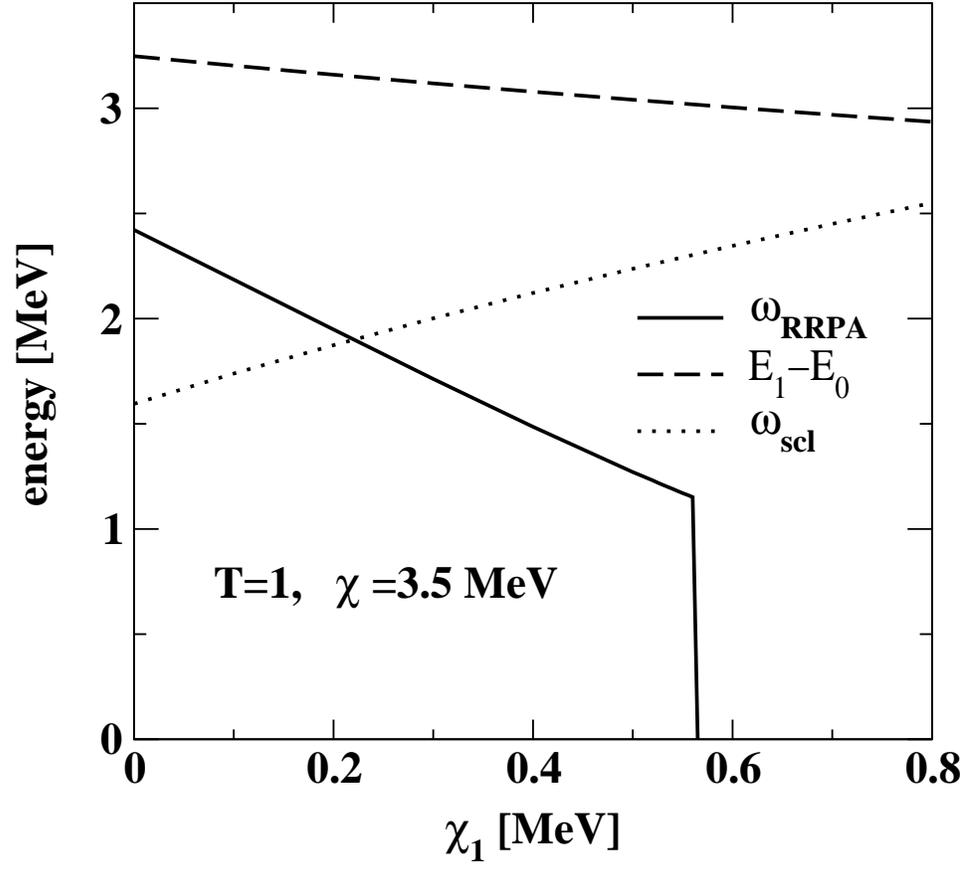,width=14cm,bbllx=0.0cm%
bblly=4cm,bburx=18cm,bbury=20cm,angle=0}}
\vspace*{-2cm}
\caption{The energies yielded by the renormalized RPA description 
(5.9)(full line), semi-classical approach (4.12) (dotted line) and by diagonalization procedure
(dashed line)  
are plotted as function of $\chi_1$ for $T=1$  
and $\chi=3.5$ MeV.}
\label{Fig 3.}
\end{figure}
\end{document}